\documentclass[flenq,10pt]{wlscirep}
\usepackage[utf8]{inputenc}
\usepackage{subfig} 
\usepackage{multirow} 
\usepackage{chemfig} 
\usepackage{xcolor} 
\usepackage{colortbl} 
\usepackage{color}
\usepackage{rotating} 
\usepackage{array,ragged2e}
\usepackage[final]{pdfpages}
\author[1,2,3,*]{M. Ropo}
\author[1,4,*]{V. Blum}
\author[1,*]{C. Baldauf}
\affil[1]{Fritz-Haber-Institut der Max-Planck-Gesellschaft, Faradayweg 4-6, D-14195 Berlin, Germany}
\affil[2]{Department of Physics, Tampere University of Technology, Finland}
\affil[3]{COMP, Department of Applied Physics, Aalto University, Finland}
\affil[4]{Department of Mechanical Engineering and Materials Science, Duke University, Durham, NC, USA}
\affil[*]{Corresponding authors: ropo@fhi-berlin.mpg.de, volker.blum@duke.edu, baldauf@fhi-berlin.mpg.de}

\title{Trends for isolated amino acids and dipeptides: Conformation, divalent ion binding, and remarkable similarity of binding to calcium and lead}

\begin{abstract}
\noindent 
We derive structural and binding energy trends for twenty amino acids, their dipeptides, and their interactions with the divalent cations Ca$^{2+}$, Ba$^{2+}$, Sr$^{2+}$, Cd$^{2+}$, Pb$^{2+}$, and Hg$^{2+}$.
The underlying data set consists of 45,892 first-principles predicted conformers with relative energies up to $\sim$4\,eV ($\sim$400\,kJ/mol). 
We show that only very few distinct backbone structures
of isolated amino acids and their dipeptides emerge as lowest-energy
conformers. 
The isolated amino acids predominantly adopt structures that involve an acidic proton shared between the carboxy and amino function. 
Dipeptides adopt one of two intramolecular-hydrogen bonded conformations C$_5$ or C$_7^{\rm{eq}}$. 
Upon complexation with a divalent cation, the accessible conformational space shrinks and intramolecular hydrogen bonding is prevented due to strong electrostatic interaction of backbone and side chain functional groups with cations. 
Clear correlations emerge from the binding energies of the six divalent ions with amino acids and dipeptides. 
Cd$^{2+}$ and Hg$^{2+}$ show the largest binding energies -- a potential correlation with their known high acute toxicities.
Ca$^{2+}$ and Pb$^{2+}$ reveal almost identical binding energies across the entire series of amino acids and dipeptides. 
This observation validates past indications that ion-mimicry of calcium and lead should play an important role in a toxicological context. 

\end{abstract}

\begin{document}
\maketitle

\section*{Introduction}
\begin{figure} 
\begin{center}
\includegraphics[width=\textwidth]{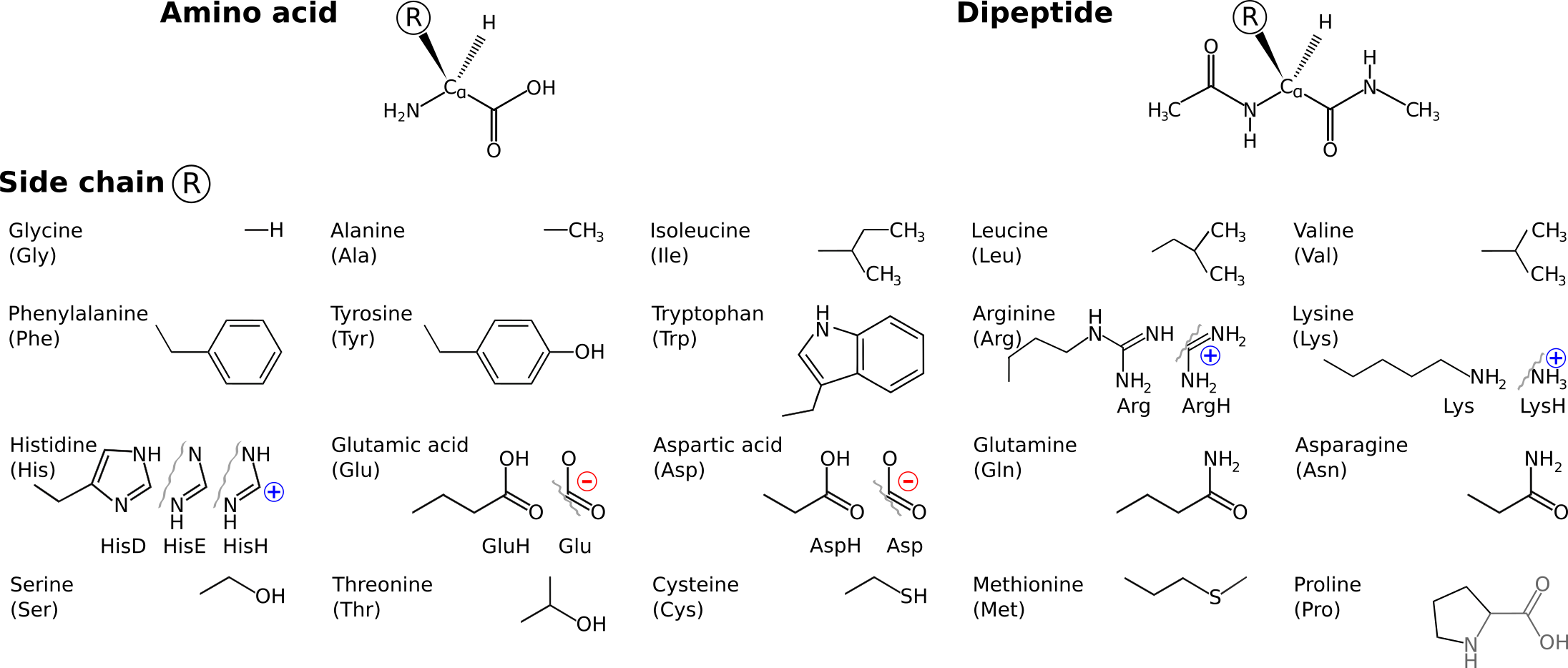} 
\end{center}
\caption{Molecular systems covered by this study. Top row:  Basic chemical formulae of an amino acid and the corresponding dipeptide. Side chains are represented by \textbf{R}. Lower panel: The chemical structures for the 20 proteinogenic side chains \textbf{R} considered in this work. Where applicable, the alternative side chain protonation states considered in this work are shown as well.}  
\label{fig:systems} 
\end{figure}

\noindent
Proteins are the machinery of life.
Their function is directly linked to their structure and dynamics.
Natural proteins are polyamides that are composed predominantly of the
twenty amino acids, shown in Figure~\ref{fig:systems}. 
Their three-dimensional structures are shaped and their dynamics are
influenced by several well-known conformational factors:  
(i)~intrinsic structural propensities of the individual building blocks, 
(ii)~intramolecular interactions such as hydrogen bonding, salt
bridges, aromatic stacking, and van der Waals interactions,  
(iii)~the surrounding medium, via bulk effects as well as by specific
interactions. 
While many of the details of protein structure arise only when
specific amino acids are combined in a chain, a first perimeter of the
available conformation space is already set out at the level of
individual amino acids.\cite{msr31_391,baldauf2012ab}   
This includes conformational preferences, e.g., through rigidity of
bond lengths and angles, or through preferred backbone conformations
defined by torsion angle patterns due to steric constraints. 
Furthermore, steric demands of side chain rotamers, electrostatics,
protonation propensity, specific chemical interactions with side
chains, and other local molecular properties are
already present at the monomer level. 

A particularly important example of specific interactions between
proteins and their environment is that with cations. About 40\% of all
proteins are known to bind metals.\cite{cr96_2239,cob3_378,jib102_1901} 
For example, Ca$^{2+}$ is essential for living organisms due to its
important role in a multitude of functions, from cell signaling to
bone growth.\cite{ebj39_825} Calcium mediated functions can be
disturbed by the presence of alternative divalent heavy metal
cations.\cite{jt132671,jib102_1901,bbrc372_341} In particular, lead
is understood to ``partially mimic the function of Ca$^{2+}$'',\cite{Florea2013}
with a range of specific, documented long-term detrimental neurotoxic
effects as a result. On the other hand, the sometimes very  
different chemical action of lead in a toxicological context compared
to Ca$^{2+}$ has also been pointed out.\cite{Simons1986} It should be
possible to establish the overall chemical 
similarity of two different ions such as Pb$^{2+}$ and Ca$^{2+}$
across a large series of potentially ligating biochemical groups based
on atomistic simulations. This task is, however, fraught with
difficulty even for simple descriptors such as binding energies. The
reason is the large space of possible molecular conformations that
must be assessed with uniform accuracy for both 
ions across a large series, even for comparatively small ligating
molecules. Without knowing what are the relevant conformers to
consider, structural trends based on total or free energies would
remain qualitative and prone to accidental omissions of relevant
conformers. Empirical potential energy surface descriptions could
certainly cover the relevant spaces, but accurate ion-molecule
interactions present significant difficulty in empirical atomistic
modeling.

In this work, we categorize the intrinsic structural properties of 
twenty proteinogenic amino acids and dipeptides, as well as their
interactions with the series of divalent cations Ca$^{2+}$, Ba$^{2+}$,
Sr$^{2+}$, Cd$^{2+}$, Pb$^{2+}$, and Hg$^{2+}$, based on a recently
published, exhaustive first-principles dataset of their possible
conformational-energy minima.\cite{db-paper,moleculedb} The complete
dataset covers 20$\times$2$\times$7=280 
molecular systems (cf. Figure~\ref{fig:systems}): 20
proteinogenic amino acid side chains attached to 
2 different backbone types, either free termini or capped
(N-terminally acetylated and C-terminally amino-methylated), in
7 distinct complexation states, i.e., either the isolated
system or in complex with one of the six cations Ca$^{2+}$, Ba$^{2+}$,
Sr$^{2+}$, Cd$^{2+}$, Pb$^{2+}$, or Hg$^{2+}$. 
Different protonation states of the side chains (basic and acidic
amino acids, see Figure~\ref{fig:systems}) or the backbone (neutral
and zwitterionic) were considered where applicable. 
The total number of conformers covered is 45,892.
This is close to the limit
of what can be accomplished computationally for a conformational
search of this extent and for the extensively benchmarked\cite{prl106_118102,cej19_11224,db-paper,jpcl1_3465,C4CP05541A,C4CP05216A,doi:10.1021/jp412055r,0953-8984-27-49-493002} level of theory used to create the database (density-functional theory (DFT) based on a van-der Waals corrected generalized gradient functional (PBE+vdW)\cite{cpc180_2175,prl77_3865,prl102_73005}, see Methods).

%\textcolor{blue}{
We here study the local, specific bonding contribution of the amino acids and dipeptides in conjunction with divalent cations, but otherwise in isolation.
This environment does, of course, not resemble biological conditions, where cations like Ca$^{2+}$ highly interact with their surrounding environment, e.g., water molecules forming hydration shells. 
However, for peptide-bound cations, such water-ion interactions do not interfere directly with the ion-peptide interactions. 
From a modeling perspective, including solvent effects either implicitly (by polarization effects)\cite{marenich_universal_2009,klamt_cosmo_2011,mennucci_polarizable_2012,ringe_function-space-based_2016} or by direct calculations of free-energy differences from molecular dynamics is in principle possible.
For the breadth of the conformational study presented here, though, such an attempt would introduce the inevitable ambiguity of \textit{which exact} solvent environment and/or which model to consider, as well as (for an explicit treatment) a sheer amount of free-energy calculations that is significantly beyond the scope of the present work. 
In the present paper, we therefore focus on the large and precisely definable total-energy contribution emerging from the specific ion-peptide bond.%}

%\textcolor{blue}{ We study the amino acids and dipeptides in conjunction with cations in the gas phase. This environment does of course not resemble biological conditions, yet is was shown experimentally that valid conclusions can be drawn from experiments in the gas phase.\cite{JarroldReview2007, Wyttenbach:2009fb, Simons_2009, arpc58_585, Schermann2007, RijsOomens15} I addition, studying biomolecules in isolation offers a clean starting point\cite{0953-8984-27-49-493002} to study and ultimately understand solvation effects by for example stepwise adding water molecules,\cite{ChutiaBlum2012} ions,\cite{cej19_11224} or other solvent-mimicking molecules.\cite{warnke_protein_2013,goth_gas-phase_2016}}

Specifically, we
show how a large-scale and systematic computational effort enables us to reveal
conformational and binding energy trends that would not be readily
apparent from isolated case studies based on different, potentially
disparate levels of theory or experimental setups. A particularly
striking example is the remarkable similarity of Ca$^{2+}$ and Pb$^{2+}$
interactions with a broad range of biologically relevant ligands,
proven below and indicating that the related
observations in toxicology\cite{Florea2013,Simons1986} emerge directly
from the fundamental underlying potential energy surface.

\section*{Results and discussion}
%\subsection{Number of conformers}
\subsection*{Trend 1: Size of conformational space} 
\begin{figure} 
\begin{center}
\includegraphics[width=1\textwidth]{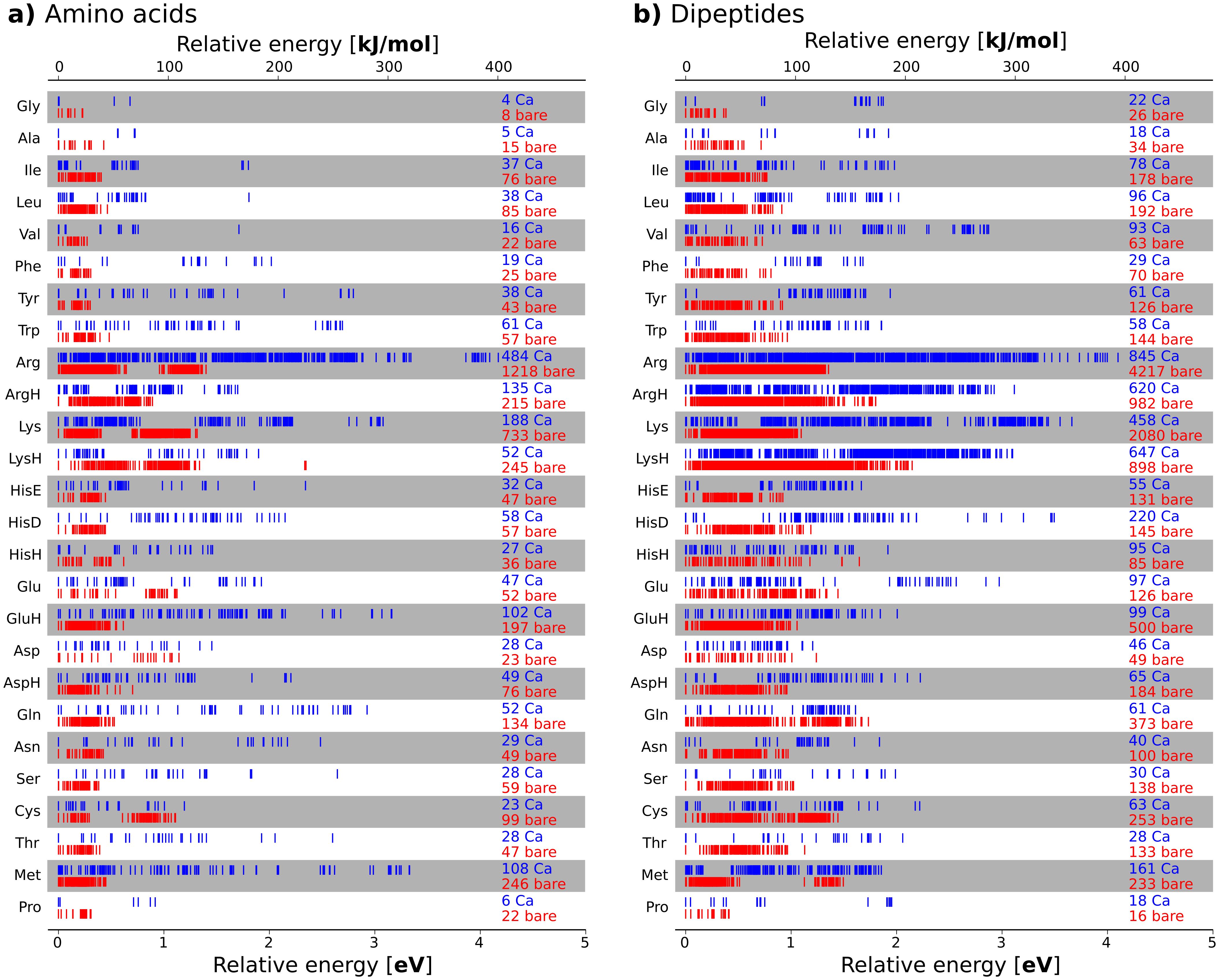}
\end{center}
\caption{The conformational hierarchies for each amino acids
  (\textbf{a})  and the capped amino acids (dipeptides) are shown for
  the isolated ("bare", red) and, alternatively, for the Ca$^{2+}$
  coordinated form (blue). The labels "Ca" and "bare" are accompanied
  by numbers that reflect the total number of conformers found for
  each system.}  
\label{fig:hierarchies} 
\end{figure}

Figure~\ref{fig:hierarchies}
summarizes the PBE+vdW conformational 
energy hierarchies and overall numbers of conformers considered in our
study. For the amino acids and dipeptides without ions, the number of
minima with the size (number of atoms) and  
flexibility (number of freely rotatable bonds) of the side chains of
the building blocks. Consequently, we predict only a few conformers
(from below ten to a few tens) for the small amino acids and
dipeptides without a side chain (Gly), with a short side chain (Ala),
or with a constrained side chain (Pro). 
In contrast, thousand(s) of conformers are predicted for the amino
acids with long and flexible side chains, especially Arg and Lys. This
number is not surprising, since the side chains alone give rise to many
different conformations that are close in energy. Tabulations of these
possible conformations, so-called rotamer
libraries,\cite{Dunbrack2002} are sometimes used in protein modeling
in order to predict, for a given backbone conformation,  a set of
probable side chain conformations. Current standard rotamer libraries
are either (i)~derived from protein 
crystal structures applying filtering and selection
criteria\cite{PROT:PROT50,rotamer} 
or (ii)~based on carefully curated sets of experimental protein
structures or model peptides (GGXGG) that were subjected to molecular
dynamics
simulations.\cite{scouras_dynameomics_2011,dyna_url} 
From the latter, rotamers can be obtained that do not carry the bias
of the crystal structures, yet they still rely on the empirical
parametrization of the underlying force fields. 
Our data set\cite{moleculedb,db-paper} offers an alternative,
empiricism free, basis for the derivation of rotamer libraries.

If the amino acids are instead coordinated to the positive ion
Ca$^{2+}$, the overall space of conformational minima contracts
significantly (1,694 and 4,103  conformers overall for the amino acids
and dipeptides, respectively). Simultaneously, the relative energy range
expands to up to about 4\,eV or 400\,kJ/mol
(Figure~\ref{fig:hierarchies}).
Evidently, the cation places a strong electrostatic constraint on the
positions of electronegative atoms and therefore reduces the accessible
conformational space.  
This effect is especially pronounced for amino acids with a flexible
side chain that interacts strongly with the cation due a lone pair, as
exemplified by the difference between the unprotonated and
protonated pairs Arg/ArgH and Lys/LysH. 
Here, the protonation results in a Coulomb repulsion between the
positively charged cation and the positively charged end group of the
amino acid side chain. 
As a results, the overall number of conformation changes from 484 to
135 for Arg/ArgH + Ca$^{2+}$ and from 188 to 52 for Lys/LysH +
Ca$^{2+}$. 

%\subsection{Conformers and isomers of uncapped amino acids}
\subsection*{Trend 2: Conformational preferences} 
\begin{figure} 
\begin{center}
\includegraphics[width=0.7\textwidth]{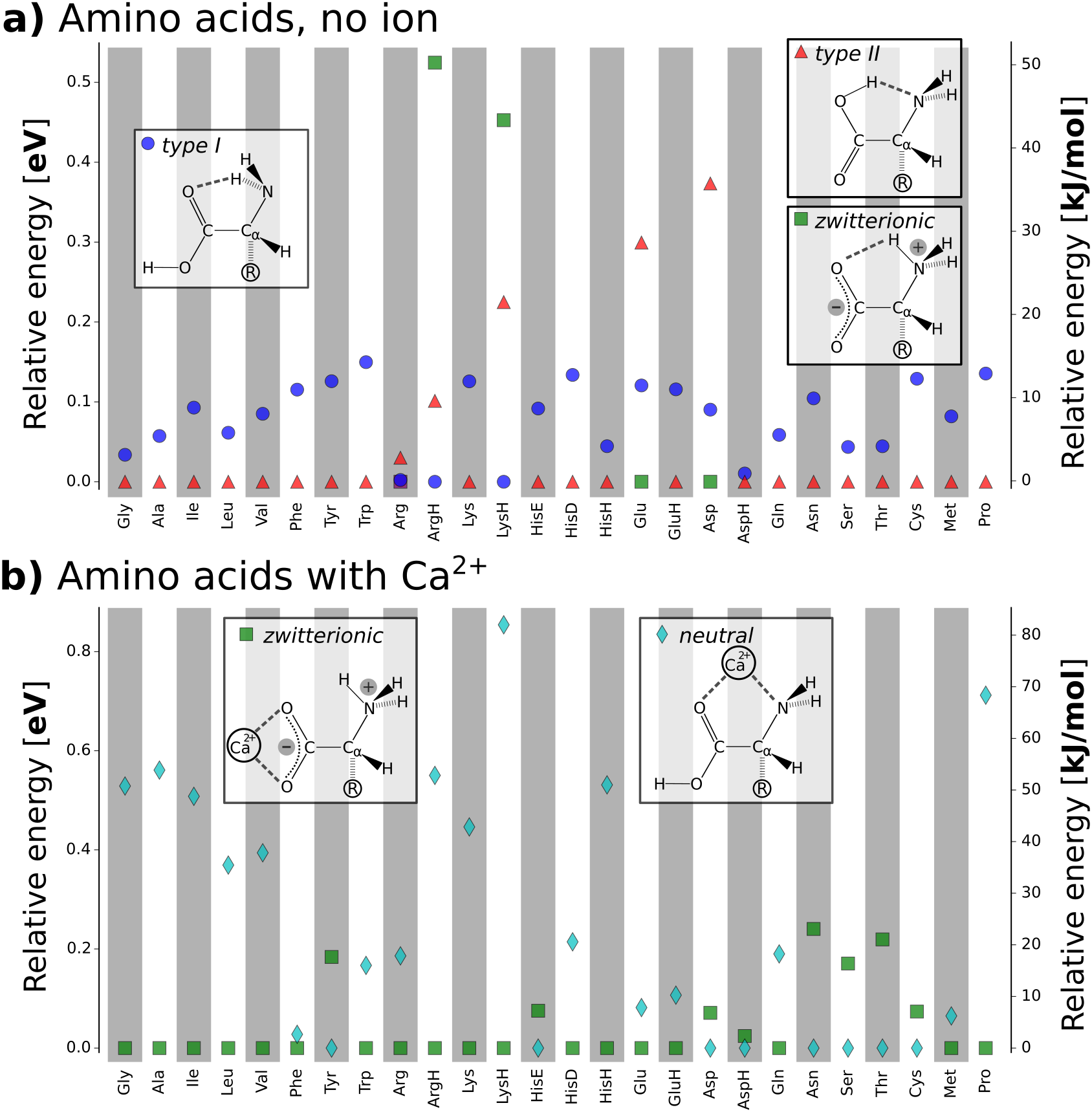} 
\end{center}
\caption{Preferred backbone conformations and protonation states for
  bare amino acids and for amino acids with
  Ca$^{2+}$. \textbf{(a)}~Schematic representations of the possible
  backbone H bonded structure types in amino acids together with a
  plot detailing the energy hierarchy of types I and II and the
  zwitterionic state for the isolated amino acids. \textbf{(b)}~The two
  basic backbone-cation conformation types for amino acids with
  Ca$^{2+}$ and a plot of their relative energies for each amino acid
  system studied. For clarity, only the lowest energy representatives
  of the respective structure types are shown. The energy of the
  respective global minimum is set to 0. The order of the amino acids
  on the $x$ axis reflects chemical groups in the following sequence:
  aliphatic, aromatic, basic, acidic, amides, alcohol/thiol, other.  
} 
\label{fig:AA_structures} 
\end{figure}

For the isolated amino
acids, the preferred conformation types are schematically summarized
in Figure~\ref{fig:AA_structures}A.
Isolated amino acids (with the exception of Arg) are found to assume
one of three basic backbone conformations (type I, type II,
zwitterionic) as their lowest-energy structure.  There is a close
relation between type II and the zwitterionic 
state: only a minuscule shift of the shared proton from the carboxylic
group to the amino function converts one type to the other. Most of
the lowest energy conformers feature the shared acidic proton between
the backbone carboxylic acid function and the backbone amino-N. Thus,
this group of lowest-energy conformers is either type II or
zwitterionic. In the case of Arg, a similar conformation is assumed by
the zwitterionic backbone carboxyl group and the side chain guanidino
function. We caution that, in several cases, the conformational
energy differences between the basic backbone conformations are rather
narrow. In these cases, changes
to the level of theory could alter the detailed hierarchy
observed.\cite{C4CP05541A,C4CP05216A,doi:10.1021/jp412055r,0953-8984-27-49-493002} 
In fact, the comparison to previous first principles studies of 
amino acids shows that different methods of calculation (level of
theory and basis set) predict different preferred
conformations.\cite{jpc100_3541,jcc28_1817} In experimental studies of
aliphatic amino acids, both states (type II/zwitterionic and type I)
have been observed,\cite{jpca102_4623,jacs127_12952} also highlighting
the close energetic competition of the different states. Nevertheless, the
emergence of just a few basic preferred backbone conformations from
the vast overall conformation space studied is remarkable and, as a
trend, robust based on the PBE+vdW level of theory applied here.

For the amino acids interacting with Ca$^{2+}$ cations, zwitterionic
and neutral/uncharged state of the backbone can be clearly
distinguished. As illustrated in Figure~\ref{fig:AA_structures}B, the
cation can either bind to the lone pairs of the amino and carboxyl
groups in the neutral/uncharged state (a.k.a. charge-solvated structure)
or interact solely with the
deprotonated and negatively charged carboxyl group in the zwitterionic
state (a.k.a. salt-bridge structure).\cite{corral_infrared_2012} 
The zwitterionic backbone state is more stable than the
uncharged backbone state for 13 of 20 amino acid-Ca$^{2+}$ systems
(see Figure~\ref{fig:AA_structures}B). 
The cation-amino acid complexes of the aliphatic amino acids as well as of Gly,
Pro, and Lys are predicted to be zwitterionic for all different
cations in this study.  
%\sout{\textcolor{red}{
%Trp and Asn prefer the uncharged/neutral backbone state when interacting with the divalent cations covered by the present study. 
%}} 
%\textcolor{blue}{
Thr and Asn prefer the uncharged/neutral backbone state when interacting with the divalent cations covered by the present study.
%} 

\begin{figure} 
\begin{center}
\includegraphics[width=0.7\textwidth]{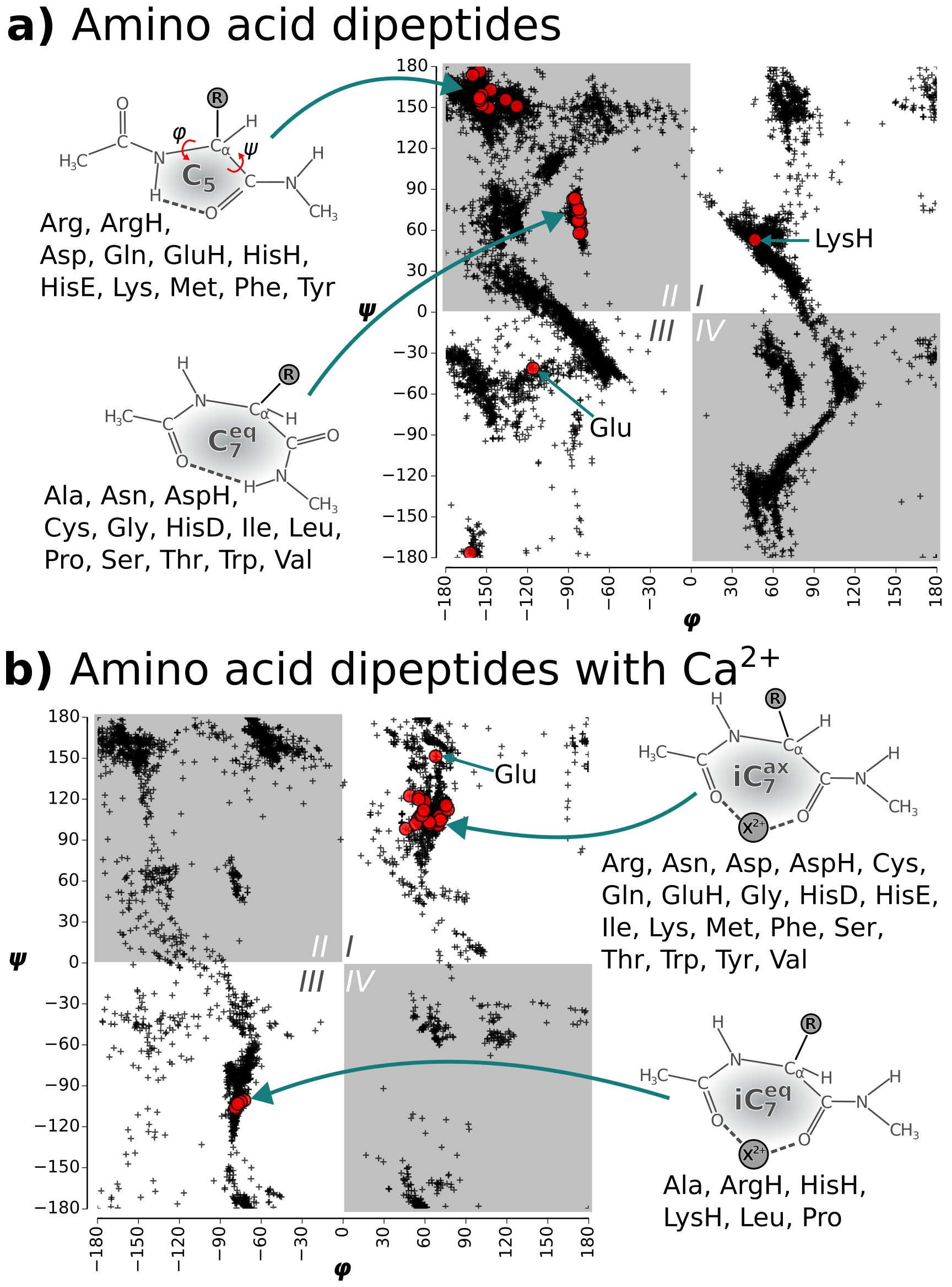} 
\end{center}
\caption{Ramachandran plots for the bare dipeptides \textbf{(a)} and
  for the dipeptides interacting with Ca${^{2+}}$ \textbf{(B)}. The
  $\phi$/$\psi$ tuples of the populated conformers of all dipeptides
  are shown by black crosses. The respective lowest energy conformers
  for each amino acid type are highlighted by red circles. Structural
  sketches illustrate the different dominant structure types of the
  global minima.
} 
\label{fig:ramachandran} 
\end{figure}

The amino-methylation and acetylation of the backbone functional
groups of the amino acids leads to the so-called dipeptides, as
schematically shown in Figures~\ref{fig:systems} and
\ref{fig:ramachandran}A. 
Since both termini resemble the local bonding environment of peptide
bond, the dipeptide form is closer to the situation of building block
embedded in a poly-peptide chain.  
In particular, end group effects such as a zwitterionic form cannot
occur in the dipeptides. 
The backbone conformational space of the dipeptides can be represented
by Ramachandran plots\cite{jmb7_95} of the torsion angles $\phi$ and
$\psi$.  
Figure~\ref{fig:ramachandran}A includes a graphical definition of both
angles. 
The two dominant conformer types found for the dipeptides are referred
to as  C$_7^{\rm{eq}}$ and C$_5$.  
The nomenclature indicates the size of the hydrogen-bonded pseudocycle
(5 or 7 members) 
The C$_7$ pseudocycle can occur in two different conformations that
are approximate mirror images. 
These images are, however, distinguished by the
\textit{\textbf{ax}ial} or \textit{\textbf{eq}uatorial} orientation of
the side chain '\textbf{R}' relative to the plane of the
hydrogen-bonded pseudocycle. 
The lowest energy conformers of each dipeptide are highlighted in the
Ramachandran plot in Figure~\ref{fig:ramachandran}A. 
The area occupied by C$_5$ is located at the 180/-180 degree border
between quadrants II and III. The area occupied by C$_7^{\rm{eq}}$ can
be found roughly in the center of quadrant II. 
All dipeptides that preferably exhibit the C$_5$ backbone conformation 
have bulky side chains. 
This is in accordance with the known propensity of bulky side chains to
enforce the formation of $\beta$ strand conformations. 
In contrast, the C$_7^{\rm{eq}}$ group almost exclusively features
members with comparatively small side chains, with the sole exception
of Trp and its large indole side chain functional group that is also
predicted to fall into the C$_7^{\rm{eq}}$ group. 

The interaction with a Ca$^{2+}$ cation has a major impact on the
predicted conformations (see Figure~\ref{fig:ramachandran}B). 
Structure types that are preferred without the cation, like C$_5$ or
C$_7^{\rm{eq}}$, are hardly populated at all in the presence of
Ca$^{2+}$.  
Instead, there are now two new dominant areas in the Ramachandran plot
in Figure~\ref{fig:ramachandran}B that differ from the preferences
exhibited by the isolated dipeptides.  
The global minima for most of the dipeptide-Ca$^{2+}$ systems are
located in quadrant I.  
The backbone oxygens of the acetyl moiety and the amino acid carbonyl
group bind the cation. 
The corresponding structure is schematically shown in
Figure~\ref{fig:ramachandran}B. 
The cation is bound to the two backbone oxygens  and thus closes an
otherwise incomplete 7-membered pseudocycle (iC$_7$). 
Similar structures were observed before by simulation and experiment
for the interaction of sodium cations with small model
peptides.\cite{cej19_11224}  
The side chain '\textbf{R}' is oriented parallel to the axis ("axial")
of this pseudocycle and is hence referred to as iC$_7^{\rm{ax}}$.   
This orientation also allows for interactions between cation and
respective side chain functional groups.  
For LysH, ArgH, HisH, Ala, and Leu, the predicted global minima are
located in quadrant III. 
The corresponding backbone structure is again characterized by a
7-membered ring that is closed by the cation. For this group of
dipeptides, the side chain is oriented equatorially, i.e. it is within
the plane of the pseudocycle. Hence, the conformation is referred to
as iC$_7^{\rm{eq}}$. 
It is the approximate mirror image of iC$_7^{\rm{ax}}$. The preference
for the equatorial side chain orientation is particularly strong for
in case of the protonated sidechains of LysH, ArgH, HisH  due to
charge repulsion. 
Finally, Pro is conformationally restricted due to the heterocycle
bridging the backbone. 

\subsection*{Trend 3: Ion binding energies} 

\begin{figure} 
\begin{center}
\includegraphics[width=0.7\textwidth]{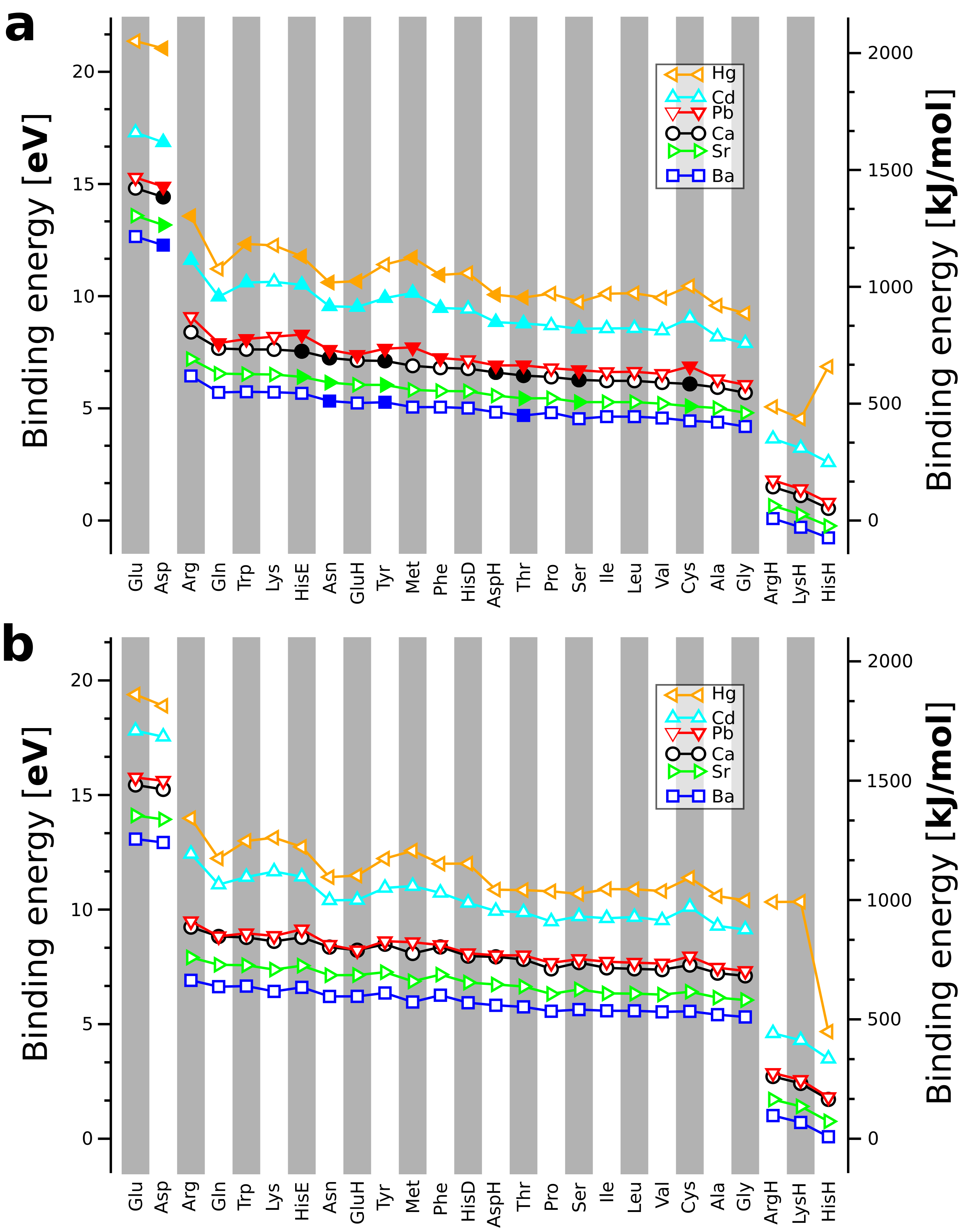} 
\end{center}
\caption{Binding affinity of the unprotected amino acids \textbf{(a)} and the dipeptides \textbf{(b)}. The building blocks are sorted according to their Ca$^{2+}$ affinity with strongest to weakest affinity from left to right. Open symbols in \textbf{(a)} indicate the zwitterionic form and filled symbols the uncharged/neutral form as the respective global minimum. The amino acids and dipeptides (and their protomers where applicable) were sorted according to the binding energy to Ca$^{2+}$ from the highest to the lowest values.} 
\label{be_norm} 
\end{figure}

In addition to the structural
effects of the cation, the binding 
energy of the individual amino acids to Ca$^{2+}$ and to other
divalent cations (Ba$^{2+}$, Cd$^{2+}$, Hg$^{2+}$, Pb$^{2+}$, or
Sr$^{2+}$) reveals distinct, remarkable trends across the series of
amino acids and their dipeptidic form.
Starting from the predicted conformers involving Ca$^{2+}$, we created
conformational energy hierarchies involving the other divalent
cations by replacing Ca$^{2+}$ with the respective alternative cation
and re-optimizing the resulting structure. 
Importantly, we include all local structure minima found with
Ca$^{2+}$, not just the Ca$^{2+}$-containing global minimum structure
for each amino acid or dipeptide. 
In this way, we can faithfully compute the binding energies for all the
alternative cations to the different proteinogenic amino acids and
dipeptides, including the various protonations states. 
We define the binding energy from total energies of the lowest-energy
conformations of the individual constituents as follows:
\begin{equation} 
E_{binding} =  E_{amino\,\,acid} + E_{cation} - E_{complex} . 
\label{Ebind}
\end{equation}

%\textcolor{red}{[REMARK: The following paragraph has been moved from the end of this section to its beginning.]}
The order of the amino acids (or dipeptides) along the $x$-axis of the
plots in Figure~\ref{be_norm} follows their affinity to the alkaline earth metals Ca$^{2+}$,
Ba$^{2+}$, and Sr$^{2+}$. 
The ranking for the amino acids binding Pb$^{2+}$, Cd$^{2+}$, and
Hg$^{2+}$ is similar but features slight deviations. 
Especially for the sulfur containing amino acids Met and Cys, the
affinities to Pb$^{2+}$, Cd$^{2+}$, and Hg$^{2+}$ increase more than
for the other amino acids (dipeptides). 
The amino acids (dipeptides) can be roughly grouped according to their
Ca$^{2+}$ affinity (from strongest to weakest): 
\begin{center}
%\sout{\textcolor{red}{
%deprotonated acids $<<$ amides, bases $<$ aromatic, acids $<$
%aliphatic $<<$ protonated bases}}\\
%\textcolor{blue}{
deprotonated acids $>>$ amides, bases $>$ aromatic, acids $>$ aliphatic $>>$ protonated bases.\\
\end{center}
%\textcolor{blue}{
Electrostatic interactions are defining here, best illustrated by the high binding energies that result from the \textit{attractive} interaction between the cation and the negatively charged deprotonated side chains of Glu and Asp and the low binding energies resulting from the \textit{repulsive} interactions between cation and positively charged side chains in case of ArgH, LysH, and HisH.
%}

Figure~\ref{be_norm} also shows the increasing affinity of the
cations to the amino acids (and dipeptides) following the order:  
\begin{center} 
%\sout{\textcolor{red}{Hg$^{2+}$ $<$ Cd$^{2+}$ $<$ Pb$^{2+}$ $\sim$ Ca$^{2+}$ $<$ Sr$^{2+}$ $<$ Ba$^{2+}$}}\\
%\textcolor{blue}{Hg$^{2+}$ (69\,pm) $>$ Cd$^{2+}$ (78\,pm) $>$ Pb$^{2+}$ (98\,pm) $\sim$ Ca$^{2+}$ (100\,pm) $>$ Sr$^{2+}$ (118\,pm) $>$ Ba$^{2+}$ (135\,pm)} 
%\textcolor{blue}{
Hg$^{2+}$  $>$ Cd$^{2+}$  $>$ Pb$^{2+}$  $\sim$ Ca$^{2+}$  $>$ Sr$^{2+}$  $>$ Ba$^{2+}$
%} 
\\(order based on average binding energies from strong to weak binding cation).
\end{center}

%\textcolor{blue}{
Importantly, the observed order of the binding energies of different cations holds uniformly across all amino acids and must therefore be intrinsic to the individual ions. 
We therefore exemplify an \textit{a posteriori} explanation of this behavior by a comparison of the lowest-energy conformations for a few representative dipeptides (Glu, Arg, GluH, and Cys) as ligands to each of the six cations covered by this study.
%}

\begin{figure}
  \begin{center}
  \includegraphics[width=\textwidth]{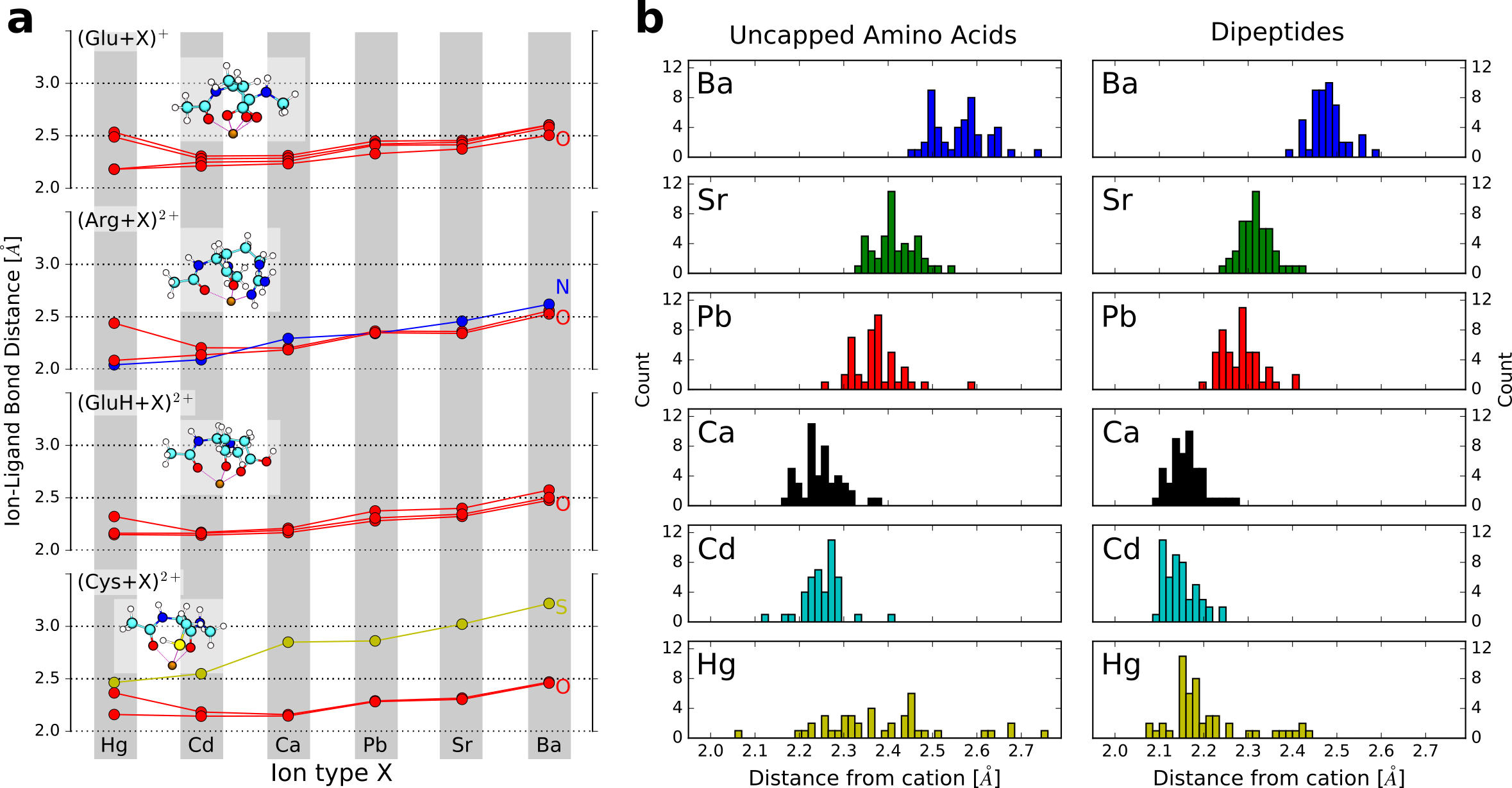}
  \end{center}
  \caption{\textbf{(a)}~Binding distances between 
  the divalent cations and their nearest ligands in the 
  lowest-energy conformations of cation-coordinated
  dipeptide forms of Glu, Arg, GluH, and Cys. The structure images
  in the insets show the Ca$^{2+}$ coordinated forms and are
  structurally equivalent for the other cations as well. Different
  ligand atoms are distinguished by different-colored curves (red: O,
  blue: N, orange: S), as noted in the figure.
  \textbf{(b)}~Histograms of cation-O distances for lowest-energy conformers over all dipeptides or uncapped amino acids and the cations covered in the study.}
  \label{fig:Distances}
\end{figure}

%\textcolor{blue}{ 
Without loss of generality, we may discuss the interaction strength in terms of distinct contributions that are well established in chemistry:
ionic (well defined if touching charged hard spheres are assumed), 
static polarizability, 
dispersion interactions, and, lastly, 
a contribution from covalent bonding that accounts for all remaining terms. We note at the outset that the computed differences in dispersion interactions between the six different ions and the peptides are much smaller than the trends shown in Figure~\ref{be_norm} and can therefore be neglected for the following discussion.
%}

%\textcolor{blue}{
In case of the alkaline earth metal cations (Ca$^{2+}$, Sr$^{2+}$, Ba$^{2+}$), the binding strength trend is well represented already by the bond distances represented in Figure~\ref{fig:Distances}a, which essentially reflect the increasing ionic radii from  Ca$^{2+}$ via Sr$^{2+}$ to Ba$^{2+}$.\cite{Shannon1976,shannon_web} For the pairs Ca$^{2+}$/Pb$^{2+}$ and Cd$^{2+}$/Hg$^{2+}$, however, ionic radii and ionic binding alone do not suffice to explain the observed trends. 
Covalent and/or polarization contributions must therefore account for the remaining differences. For a more quantitative description in terms of our own data, we plot the ion-ligand binding distances for all six cations considered to Glu, Arg, GluH, and Cys in Figure~\ref{fig:Distances}a. 
%}

%\textcolor{blue}{
Pb$^{2+}$ is in principle larger than Ca$^{2+}$, consistent with our data. By purely ionic considerations, Pb$^{2+}$ should therefore bind somewhat less strongly than Ca$^{2+}$. However, Pb$^{2+}$ features a relatively shallow filled $s$ shell in Pb$^{2+}$ that is absent in Ca$^{2+}$. Thus, Pb$^{2+}$ should be slightly more polarizable and may have a slightly larger covalent contribution to the binding strength, which results in the comparable binding energy trend for Pb$^{2+}$ and Ca$^{2+}$.
%}
%The fact that Pb$^{2+}$ binds almost exactly like Ca$^{2+}$ is likely explained by the presence of its additional filled $s$ valence shell.}

%\textcolor{blue}{
What sets Hg$^{2+}$ and Cd$^{2+}$ apart from Ca$^{2+}$, Sr$^{2+}$ and Ba$^{2+}$ is a relatively shallow filled $d$ shell for Hg$^{2+}$/Cd$^{2+}$. Ionic, static polarizability and/or residual covalent contributions should thus all lead to an overall stronger binding of Hg$^{2+}$ and Cd$^{2+}$ to the dipeptides than for the alkaline earth metals, as we observe.
Between Hg$^{2+}$ and Cd$^{2+}$, however, Hg$^{2+}$ is larger than Cd$^{2+}$ in terms of tabulated ionic radii for the same coordination number,\cite{Shannon1976,shannon_web} and thus Cd$^{2+}$ should bind more strongly for otherwise equivalent conformation. Yet, the opposite is the case in Figure~\ref{be_norm}, \textit{i.e.}, Hg$^{2+}$ binds significantly more strongly. The reason can be discerned from Figure~\ref{fig:Distances}a. While Cd$^{2+}$ retains approximately equal distances from its ligands, Hg$^{2+}$ changes its local coordination shell to pull two of the ligands closer than the others, in fact closer than any of the ligands of Cd$^{2+}$. This suggests a strongly covalent and/or static polarizability driven contribution to the Hg$^{2+}$-dipeptide bond. The orbital-based explanation likely resides in the fact that the 5$d$ shells of Hg$^{2+}$ are shallower than the 4$d$ shells of Cd$^{2+}$. As a result, the frontier orbitals of Hg$^{2+}$ are more flexible and have a larger incentive to participate in some form of residual bond. Overall, Figure~\ref{fig:Distances}a reveals that the coordination chemistry of Hg$^{2+}$ is in detail different from that of Cd$^{2+}$, explaining the much stronger overall binding strength of Hg$^{2+}$ to the amino acid and dipeptide ligands studied here.
%}

%\textcolor{blue}{
The trend that we just discussed for a few selected amino acid dipeptides holds over the whole range of amino acids and dipeptides as is illustrated by plots in the Supplementary Information similar to the ones shown in Figure~\ref{fig:Distances}a.
Additionally, Figure~\ref{fig:Distances}b shows histograms of cation-oxygen distances across the respective lowest-energy structures of all amino acids and dipeptides considered in this study. The histograms also reflect the trends of Figure~\ref{fig:Distances}a, e.g. the increase of the median distances for Ca$^{2+}$, Sr$^{2+}$, and Ba$^{2+}$ or the change of the one-peak distribution of Cd$^{2+}$ to a multi-modal distribution for Hg$^{2+}$ that indicates strongly covalent and/or static polarizability driven cation-O interactions existing alongside purely ionic interactions for Hg$^{2+}$.
%}

% VB: I would leave the following paragraph, now commented, away. There is some explaining to do to equate the 2-fold Hg shannon radius with the 4-fold Cd radius, and anyone teaching general chemistry may still grumble that Hg is, in general, larger. But I do not think we need the Shannon radius at all ... this would eliminate a risk.

%\textcolor{blue}{
%Essentially the same relation becomes obvious if we compare the %binding energy trend to Shannon's ionic %radii.\cite{Shannon1976,shannon_web}
%We chose for each ion the lowest tabulated ionic radius associated %with a charge of +2 
%(Hg$^{2+}$:69\,pm; Cd$^{2+}$: 78\,pm; Pb$^{2+}$: 98\,pm; Ca$^{2+}$: %100\,pm; Sr$^{2+}$: 118\,pm; Ba$^{2+}$: 135\,pm).
%The cation with the smallest Shannon ionic radius (Hg$^{2+}$) has %the highest affinity to the amino acids and dipeptides and the trend %continues: with increasing ionic radius, the binding energy %decreases.
%For the smaller cations the electrostatic contributions are highest %due to them being closer to the interaction partners.}

\subsection*{Trend 4: Ion mimicry and toxicity} 
We next attempt to correlate
toxicological effects of the different ions with their respective
interactions across the series of amino acids and dipeptides. For
acute (short-term) toxicity, we consider median lethal doses 
(LD$_{50}$)\cite{ld50} for the
chloride salts (selected because of their generally high solubility)
of the respective divalent cations. 
The LD$_{50}$ is the dose of a toxin required to kill half of the
individuals in a test population; it is thus a measure of acute, not
long-term toxicity. An interesting trend emerges when comparing these
LD$_{50}$ values to the binding energy trends reported in
Figure~\ref{be_norm} of the present study. Ca$^{2+}$ might naturally
serve as a reference point in this comparison. The chloride salt is
moderately toxic with an LD$_{50}$ of 2,301\,mg/kg. Sr$^{2+}$ exhibits
relatively low binding energies to the amino acids. The LD$_{50}$ of
the respective chloride is also moderate with 1,253\,mg/kg. Toxic
effects of Strontium result from radioactive isotopes and not from the
binding competition of Sr$^{2+}$ with Ca$^{2+}$. Pb$^{2+}$ and
Ca$^{2+}$ show almost exactly the same binding energy trend. Their acute
toxicities are similar and not high, with LD$_{50}$ values of
1,947\,mg/kg and 2,301\,mg/kg, respectively, for PbCl$_2$ and
CaCl$_2$. The divalent cations of mercury and cadmium show the highest
binding energies in our comparison. Interestingly, the acute
toxicities of CdCl$_2$ and HgCl$_2$ are also much higher than those of
SrCl$_2$, CaCl$_2$ and PbCl$_2$, with LD$_{50}$ values of values of
107 and 47\,mg/kg for Cd and Hg, respectively. Indeed, cadmium ions
bind tightly to proteins and compete with Zn$^{2+}$ and Ca$^{2+}$ and
especially also with sulfur bound Cu$^+$ and with iron in iron-sulfur
centers in proteins.\cite{Maret2013} 
%The mechanism of toxicology of
%mercury is complex due to the
%multiple forms in which mercury can
%be present. 
%The main effect of the divalent Hg$^{2+}$ cations are protein precipitation and enzyme inhibition, both apparently linked to interaction with thiol groups.\cite{Carvalho02052008,Broussard2002,Keil01122011}
For the cations Hg$^{2+}$, Cd$^{2+}$, Pb$^{2+}$, Ca$^{2+}$, and
Sr$^{2+}$, we might thus infer a tentative correlation of acute
toxicities with binding energy trends. A clear outlier, however, is
Ba$^{2+}$. Figure~\ref{be_norm} shows it to be the cation in this
study with the lowest affinity to the amino acids and
dipeptides. In contrast, its known acute toxicity is rather high, with
LD$_{50}$ values ranging from 100 to 300\,mg/kg. Indeed, 
the toxic species is the bare cation and not (as in the case of Hg or
Cd) the protein bound cation. Due to its large size and weak
binding energy for a divalent cation, Ba$^{2+}$ may interfere with K$^+$
channels instead and consequently has effects on, e.g., the function
of muscle tissue.\cite{welsh1983,Raisbeck} 

We finally return to the remarkable, almost exact quantitative
agreement of the computed binding energies for Ca$^{2+}$
and Pb$^{2+}$, respectively, in Figure~\ref{be_norm}. This trend is
here revealed across an entire \textit{series} of molecular
ligands that are central to biochemistry. Consistent with low acute
toxicities, their similarity 
supports directly the available evidence that Pb$^{2+}$ can mimic
Ca$^{2+}$:\cite{Florea2013} individual amino side group functions will
not distinguish these two ions. Additionally, tabulated qualitative
descriptors such as their standard hydration enthalpies (Pb$^{2+}$:
$-$1479.9 kJ/mol, Ca$^{2+}$: 1592.4 kJ/mol)\cite{Burgess1978} and
Shannon ionic radii (octahedral: Pb$^{2+}$: 98\,pm, Ca$^{2+}$: 100\,pm)\cite{Shannon1976,shannon_web} 
are also close. Yet, Pb and Ca are certainly not
chemically identical,\cite{Simons1986} e.g., due to possible
multivalency of Pb. Thus, Pb$^{2+}$ may fly ``under the radar'' of
typical factors distinguishing ions, helping facilitate its more
subtle (neuro)toxic actions\cite{Florea2013} or other detrimental
effects on protein folding, maturation, and interaction in some
instances.\cite{Garza2006} 

\section*{Summary}
This study is a first step towards an unbiased understanding of peptide cation interactions from first principles. 
%\textcolor{red}{\sout{By considering only a single amino acid/dipeptide we do consider steric effects only to a limited extend.
%Next steps could include the reconstruction of protein cation binding sites in a similar fashion a outlined in the review articles by Dudev and Lim and in references cited therein.}}
%\textcolor{red}{\sout{Nevertheless, a}} \textcolor{blue}{A} 
A large, first-principles data base of more than 45,000
conformers of 20 proteinogetic aminoacids, their dipeptides, and their
coordination with the six divalent cations Ca$^{2+}$, Ba$^{2+}$,
Sr$^{2+}$, Cd$^{2+}$, Hg$^{2+}$, and
Pb$^{2+}$\cite{db-paper,moleculedb} allows us to identify trends across
an entire series of biochemically relevant functionalities:  
\begin{itemize}
\item In the gas phase and at the  PBE+vdW level of theory, the
  type~II/zwitterionic form of the amino acid backbone dominates over
  type~I. 
\item The dipeptides (acetylated and amino-methylated amino acids)
  assume only two distinct types of low-energy conformations, C$_5$
  and C$_7^{\rm{eq}}$.%\textcolor{blue}{???Should there be more
                      %references here?}%In agreement with trends
                      %reported earlier in the literature (see for
                      %example Yuan \textit{et al.}\cite{jpca118_7876}
                      %article and its references)  
\item The conformational space of amino acids and dipeptides, as
  measured by the number of local minima found for each system,
  contracts upon coordination with divalent cations (with Ca$^{2+}$
  attachment, significantly fewer minima are found). Their
  conformational freedom is reduced by strong electrostatic
  interactions of backbone and side chain functionalities with the
  cation.   
\item The interaction of amino acids and Ca$^{2+}$ occurs preferably
  via the deprotonated (negatively charged) backbone carbonyl
  functions. In the cases of the dipeptides, the Ca$^{2+}$ cation
  preferentially interacts with the backbone carbonyl groups of the
  amino acid and the acetyl capping. 
\item The heavy-metal cations Hg$^{2+}$ and Cd$^{2+}$ bind more
  strongly to the amino acids and dipeptides than
  Ca$^{2+}$. 
\item We can construct an {\it a posteriori} plausible correlation
  between the general binding energies of divalent cations to the
  amino acids and the acute toxicities of their chloride salts. The
  strong binding energies of Cd$^{2+}$ and Hg$^{2+}$ correlate with
  their much higher acute toxicities, while the weakest binder,
  Ba$^{2+}$, is known to affect the function of another much more
  weakly binding cation, K$^{+}$. 
\item At the PBE+vdW level of theory, Pb$^{2+}$ shows binding energies
  to all studied amino acids and dipeptides that are virtually
  identical to those of Ca$^{2+}$. Pb$^{2+}$ thus has the ability to
  mimic much of the function of Ca$^{2+}$, but not precisely,
  consistent with its interference with Ca$^{2+}$ functions over
  longer time scales and/or in specific circumstances. 
\end{itemize}

The insights generated in this work are inherently enabled by having access to an essentially exhaustive conformational energy data set for a wide range of amino acids, their dipeptides, and their various cation bound forms. 
%\textcolor{blue}{
The focus of our study is to study general trends across a large portion of chemical space. 
This of course comes at the expense of investigating more subtle effects, e.g. effects of electronic 
correlation energy treatments at higher levels and the effects of finite temperature, and entropy
are not reflected but could be included in future refinements of the existing
database.
%\textcolor{blue}{
By considering only single amino acids/dipeptides we consider steric effects only to a limited extend. 
Next steps could include the reconstruction of protein cation binding sites in a similar fashion a outlined in the review articles by Dudev and Lim and in references cited therein.\cite{dudev_principles_2003,dudev_metal_2008,dudev_competition_2014}
%}
%\textcolor{blue}{
%In the realm of biomolecules, the solvent environment, which we neglect in this study, is of course very important. A promising route is for example the re-evaluation of general trends, including a structure refinement, by incorporating solvent effects by means of continuum models.\cite{marenich_universal_2009,klamt_cosmo_2011,mennucci_polarizable_2012,ringe_function-space-based_2016} The effects of direct interactions between solvent molecules and peptide-cation systems could for example be studied by \textit{ab initio} molecular dynamics simulation involving explicit water molecules.\cite{cej19_11224,ChutiaBlum2012}
In addition and as noted in the introduction, the environment or solvent medium surrounding cations bound to peptides is, of course, very important. For a quantitative understanding of specific environmental conditions, we believe that dedicated simulations incorporating effective continuum models\cite{marenich_universal_2009,klamt_cosmo_2011,mennucci_polarizable_2012,ringe_function-space-based_2016} or simulations of free energy differences by explicit molecular dynamics constitute excellent follow-up opportunities to our study.
%} 

% VB: I am not sure the following sentence is still needed. It's kind of said above already.

%These and other effects will most certainly make a
%difference when analyzing the detailed aspects of peptide and protein
%conformations on specific processes in biology and
%chemistry. 

Nonetheless, the emergence of unambiguous overall trends
from an entirely non-empirical first-principles treatment (not relying
on any specific ``chemical intuition'' at the outset) in our study is
encouraging, showing the relevance of trends at the Born-Oppenheimer
potential energy surface as a basic quantity and highlighting the
power of systematically scanning large segments of chemical space with
high computational accuracy to underpin empirically suspected
trends.  

\section*{Methods}
\noindent
The level of theory employed to assemble the conformer database\cite{db-paper} 
is a high-accuracy\cite{Lejaeghereaad2016} numeric atom-centered basis set
implementation\cite{cpc180_2175,Havu2009} of density-functional theory (DFT) 
in the Perdew-Burke-Ernzerhof (PBE)\cite{prl77_3865}
generalized-gradient approximation, combined with the pairwise van der Waals
correction by Tkatchenko and Scheffler (PBE+vdW).\cite{prl102_73005}
The reliability of PBE+vdW for the peptide structure problem has been
established by (i)~comparisons to CCSD(T) benchmark investigations for
oligo-alanine systems\cite{prl106_118102} and alanine dipeptides with
Li$^+$\cite{cej19_11224}, (ii)~comparison to carefully performed
basis-set extrapolated MP2 calculations\cite{db-paper}, and
comparisons to experimental spectroscopic benchmarks.\cite{jpcl1_3465}
Importantly, the actual \textit{conformations} identified by PBE+vdW
remain meaningful even in cases where, for larger oligopeptides,
more expensive higher levels of theory are needed to reproduce small
energetic differences to the exact conformational energy
\textit{hierarchy} found 
in experiment.\cite{C4CP05541A,C4CP05216A,doi:10.1021/jp412055r,0953-8984-27-49-493002} 

As described in detail in Ref. \cite{db-paper},
for each of the amino acids and dipeptides with and without
Ca$^{2+}$, a global basin-hopping search with
\textsc{Tinker}\cite{jcc87_1016,jpcb107_5933} was performed using the
OPLS-AA force field\cite{jacs118_11225} to pre-screen for relevant conformations. 
This set of conformations was then relaxed at the PBE+vdW level. After this
global search step, a local conformational refinement was 
performed by PBE+vdW based replica-exchange molecular dynamics
(REMD)\cite{prl57_2607,cpl314_141,C1FD00027F} followed again by
geometry relaxations with PBE+vdW. 
For a subset of seven dipeptides (Ala, Gly, Phe, Val, Ile, Trp, Leu)
and by comparing to an independently performed genetic algorithm
search at the same level of theory,\cite{doi:10.1021/acs.jcim.5b00243}
this search was shown to be essentially complete.
Conformers for amino acids and dipeptides in complex with Ba$^{2+}$,
Sr$^{2+}$, Cd$^{2+}$, Pb$^{2+}$, and Hg$^{2+}$ were generated by
substituting a different ion into the geometries of the Ca$^{2+}$
complexes and locally optimizing the new geometry.

%\bibliography{aa_metal}

\section*{Acknowledgements}
The authors thank Matthias Scheffler (FHI Berlin) for continuous support of this work.

\section*{Author contributions}
M.R., V.B., C.B. designed the study.
M.R. performed the simulations.
M.R., C.B. analyzed the data.
M.R., V.B., C.B. wrote the article.

\section*{Competing financial interests}
The authors declare no competing financial interests.

\newpage
\section*{Supplementary Information}
\noindent
Binding distances between the divalent cations and their nearest ligands in the lowest-energy conformations of 20 cation-coordinated proteinogenic dipeptides. Different ligand atoms are distinguished by different-colored curves (red: O, blue: N, orange: S), as noted in the following figures.

%\includepdf[scale=0.85,pages=-,pagecommand={}]{PDFs/2009_Baldauf-JThrombHaemost.pdf}
\includepdf[scale=0.65,pages=-]{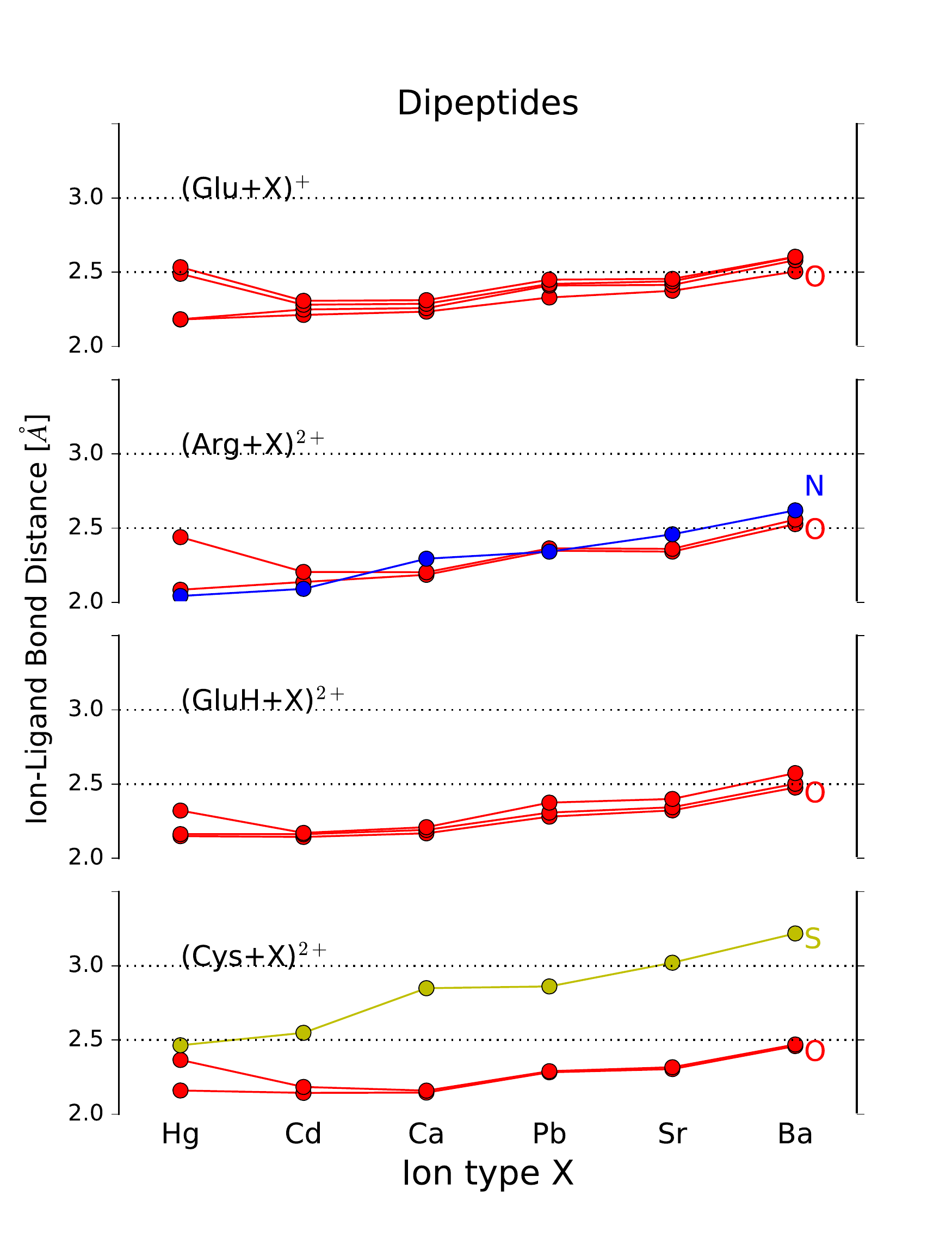}

\end{document}